\documentclass[12pt]{article}
\usepackage{amsmath,amssymb}    
\usepackage{amsfonts}
\usepackage{enumerate}
\usepackage{ytableau}
\usepackage{hyperref}
\usepackage{bm}
\usepackage{tikz}
\usepackage{blkarray}  
\begin{document}
\def\be{\begin{equation}}
\def\bea{\begin{eqnarray}}
\def\ee{\end{equation}}
\def\eea{\end{eqnarray}}
\def\d{\partial}
\def\eps{\varepsilon}
\def\la{\lambda}
\def\b{\bigskip}
\def\nn{\nonumber \\}
\def\p{\partial}
\def\t{\tilde}
\def\h{{1\over 2}}
\def\be{\begin{equation}}
\def\bea{\begin{eqnarray}}
\def\ee{\end{equation}}
\def\eea{\end{eqnarray}}
\def\b{\bigskip}
\def\u{\uparrow}
\def \byt{\begin{ytableau}}
\def \eyt{\end{ytableau}}
\newcommand{\tabincell}[2]{\begin{tabular}{@{}#1@{}}#2\end{tabular}}
\def\be#1\ee{\begin{equation}\begin{split}#1\end{split}\end{equation}}
\def\({\left(}
\def\){\right)}
\def\[{\left[}
\def\]{\right]}

\def\ta{\theta}
\def\p{\partial}
\def\s#1{|_{#1}}

\makeatletter
\def\blfootnote{\xdef\@thefnmark{}\@footnotetext}  
\makeatother


\title{\textbf{Asymmetric $\lambda$--deformed cosets}}

\vspace{14mm}
\author{
 Jia Tian$^{1,3}$, Jue Hou$^{1}$,and Bin Chen$^{1,2,3}$\footnote{bchen01, houjue, wukongjiaozi@pku.edu.cn}
}
\date{}

\maketitle

\begin{center}
{\it
$^{1}$Department of Physics and State Key Laboratory of Nuclear Physics and Technology,\\Peking University, No.5 Yiheyuan Rd, Beijing 100871, P.~R.~China\\
\vspace{2mm}
$^{2}$Collaborative Innovation Center of Quantum Matter, No.5 Yiheyuan Rd, Beijing 100871, P.~R.~China\\
$^{3}$Center for High Energy Physics, Peking University, No.5 Yiheyuan Rd, Beijing 100871, P.~R.~China
}
\vspace{10mm}

\end{center}

\begin{abstract}
We study the integrable asymmetric $\lambda$-deformations  of the $SO(n+1)/SO(n)$ coset models, following the prescription proposed in \cite{AsyLambda}. We construct all corresponding deformed geometries  in an inductive way. Remarkably 
we find a $Z_2$ transformation which maps the asymmetric $\lambda$--deformed  models to the symmetric $\lambda$--deformed models.

\end{abstract}

\baselineskip 18pt

\newpage



\section{Introduction}
\label{Intorduction}
Integrability plays a key role in obtaining exact results in field theories and string theory. After witnessing a remarkable progress in understanding the integrable string theories it is clearly interesting and important  to construct new integrable models. In the last ten years, based on sigma models on group or coset manifolds powerful tools  such as $\eta$-- \cite{BosonicYangBaxter,DelducEta,eta} and $\lambda$--deformations \cite{SftsGr,MoreLmb,DSTLmbAdS5} have been developed to investigate this issue. 

The original $\lambda$--deformation \cite{SftsGr} gives an interpolation between the (gauged) Wess-Zumino-Witten (WZW) model \cite{WZW}and the non-Abelian T-dual of the Principle Chiral model (PCM) \cite{PCM}. The deformations of $AdS_p\times S^p$ have been successfully constructed in \cite{MoreLmb,DSTLmbAdS5}. The novel idea behind the construction of $\lambda$--deformation is to combine different integrable models through a gauging procedure.  Following the same idea, various of generalizations of $\lambda$--deformation have been proposed \cite{GenLambda1,GenLambda1.5,GenLambda2}.  Recently, the authors in \cite{AsyLambda} introduced a new generalization named asymmetric $\lambda$-deformation by modifying the gauging procedure. The key observation in \cite{AsyLambda} is that different choices of anomaly free gauges can be made in the deformation. It is somehow similar to the gauged WZW model with $U(1)$ symmetry, in which case one can choose  either the vector or axial gauge and the  two resulted gauged theories are T-dual to each other \cite{Duality}. Now the deformation breaks the axial-vector duality  since the deformation destroys the isometries of the background. To have a non-trivial asymmetric $\lambda$--deformation, the starting model has to possess a Lie algebra with non-trivial outer automorphism group. 

In this article, we will study the asymmetric $\lambda$-deformation of  the $SO(n+1)/SO(n)$ coset models, and pay special attention to the case that  $n$ is odd and the corresponding Lie algebra admits a $Z_2$ outer automorphism group.
A physical motivation to study the deformation of this class of coset models is that the coset $SO(n+1)/SO(n)$ is isomorphic to $S^{n}$ and by an analytic continuation they can be transformed into $AdS_{n}$. Then it would be possible to embed the deformed models into supergravity.

The paper has the following organization: In section 2, we review the construction of asymmetric $\lambda$--deformation introduced in \cite{AsyLambda}. To apply the construction, we need to find a suitable coset representative and explicit forms of the outer automorphism in this given representative. So in section 3, we first find the outer automorphism of the $SO(n)$ group following \cite{Ganor:2002ya} and then develop a gauge fixing scheme similar to \cite{Grigoriev:2007bu}. After constructing the deformed models we prove the existence of a $Z_2$ transformation which maps asymmetric $\lambda$--deformed geometry to the symmetric deformed ones. As a by product, we provide a simple recursive method to construct the deformed geometries in our choice of coset representative. Some technical details are presented in the appendices.\par

\section{Asymmetric $\lambda$-deformation}
\label{AsymmetricLambda}
\renewcommand{\theequation}{2.\arabic{equation}}
\setcounter{equation}{0}

In this section, we  briefly review the integrable  asymmetric $\lambda$-deformation. Here we only focus on  symmetric coset models. Similar constructions can also be applied  to group and supercoset models. For the details  we refer the original article \cite{AsyLambda}. 
To introduce the deformation of a coset model $G/H$, we begin with separating the generators $T^A$ of the group $G$ into $T^a$ and $T^\alpha$ corresponding to the subgroup $H$ and the coset $G/H$  respectively \footnote{Here we follow the convention in \cite{SftsGr}, the Latin index ($a$)  denotes the component in the subsgroup and Greek index ($\alpha$) denotes components in the coset $G/H$.}, and then defining the Maurer-Cartan forms
\bea\label{MCforms} 
&&L^A_\mu=-i \mbox{Tr}(T^A g^{-1}\p_\mu g),\quad R^A_\mu=-i \mbox{Tr}(T^A \p_\mu g g^{-1}),\nn
&& R_\mu^A=D_{AB}L^B_\mu,\quad D_{AB}=\mbox{Tr}(T_A g T_B g^{-1}),\quad g\in G.
\eea 
The asymmetric $\lambda$--deformation of coset model $G/H$ is constructed by performing three steps:
\begin{enumerate}[(i)]\label{Steps}
	\item Combine $S_{PCM}(\hat{g})$ on coset $G/H$ with the $S_{WZW,k}(g)$ on the same group $G$, where 
	\bea
&&	S_{\mathrm{PCM}}\left(\hat{g}\right)=-\frac{\kappa^2}{\pi} \int \operatorname{Tr}\left(\hat{g}^{-1} \partial_{+} \hat{g} \hat{g}^{-1} \partial_{-} \hat{g}\right),\nn
&&S_{\mathrm{WZW},k}(g)=-\frac{k}{2 \pi} \int \text{Tr}\left(g^{-1} \partial_{+} g g^{-1} \partial_{-} g\right)+\frac{ k}{6 \pi} \int_{B} \text{Tr}\left(g^{-1} d g\right)^{3}.
	\eea
	
	\item Gauge the group $G$ whose $G_L$ action is given by
	\bea \label{GaugeTrans}
	g\rightarrow g_0^{-1}g \tilde{g}_0,\quad \hat{g}\rightarrow g_0^{-1}\hat{g},
	\eea 
	where $g_0=exp(G^A T_A)\in G$ and $\tilde{g}_0=exp(G^A \tilde{T}_A)\in G$ have the same parameters $G^A$ but they are generated by different embeddings $T_A$ and $\tilde{T}_A$ of the subalgebra. These two embeddings are related by  a linear transformation $W$, such that $\tilde{T}_A=W(T_A)=W^B{}_A T_B$. To avoid the gauge symmetry anomaly  the transformation $W$ has to be a metric-preserving automorphism of the Lie algebra, i.e.,
	\be
	W([T_A,T_B])=[W(T_A),W(T_B)],\quad \text{Tr} (W(T_A)W(T_B)) = \text{Tr}(T_A T_B).
	\ee
	Non-equivalent choice of $W$ is characterized by the outer automorphism group of the Lie algebra. In particular, the choice $W=I$ which is usually called the vector gauge  leads to the standard $\lambda$--deformation \cite{SftsGr}. Therefore only when $W$ is a non-trivial element of  the outer automorphism group, following \cite{AsyLambda} we will call it axial gauge,  the deformation can be potentially non-trivial. 
\item Integrate out the gauge field and fix the gauge $\hat{g}=I$.	
\end{enumerate}
These procedures give the final action of deformed model:
\be\label{DeformedAction}
S_\lambda(g,W)&=S_{WZW,k}(g)-\frac{k}{\pi}\int L_-\(D-\Omega^T W^{-1}\)^{-1}R_+\\
&=\frac{k}{2\pi}\int L_-\[1-2\(D-\Omega^T W^{-1}\)^{-1}D\]L_+,\\
\ee
where $L,R$ and $D$ are quantities defined in \eqref{MCforms} and the operator $\Omega$ is given by
\bea \label{Omega}
\Omega(T_A)=\Omega^B{}_{A}T_B,\quad \Omega^B{}_{A}=\left(
\begin{array}{cc}
	I_{ba} & 0 \\
	0 & \lambda^{-1}I_{\beta\alpha} \\
\end{array}
\right),\quad 0<\lambda<1.
\eea 

The background geometry of the target space of this model is given by\footnote{The Kalb-Ramond field vanishes for a similar argument given in \cite{Grigoriev:2007bu,GenLambda1}} ,
\bea\label{Metric}
&&ds^2=\frac{k}{2\pi} L^T\[1-MD-(MD)^T\]L,
\eea
with 
\bea
&& M\equiv\(D-\Omega^T W^{-1}\)^{-1},\nn
&&e^{-2\Phi}=e^{-2\Phi_0}\mbox{det}(D W-\Omega).
\eea
With the help of the identity $DD^T=1$ and\footnote{$W$ can always be chosen to be diagonal.} $WW^T=1$,  the metric can be simplified as 
\be
ds^2&=\frac{k}{2\pi} L^T\[1-MD-(MD)^T\]L\\
&=\frac{k}{2\pi} L^T M\[M^{-1}M^{-T}-DM^{-T}-M^{-1}D^T\]M^{T}L\\
&=\frac{k}{2\pi} (M^T L)^T \[\Omega^T\Omega-1\]M^{T}L\\
\ee
Substituting \eqref{Omega} and using the inversion formula of a block matrix, the metric can be cast into the form
\bea \label{FinalMetric}
&&ds^2=\frac{k}{2\pi}\frac{1-\lambda^2}{\lambda^2}e^T P^TP e,~~
\eea
where
\bea
e_\alpha&=&L_\alpha-D^T_{a\alpha }(D_{ab}-W_{ab})^{-T}L_b,\nn 
 (P^{-T})_{\alpha\beta}&=&\left[D_{\alpha\beta}-D_{\alpha b}(D_{ab}-W_{ab})^{-1}D_{a\beta}-\frac{1}{\lambda}W_{\alpha\beta}\right].
\eea 
Here $e$ are the frames of the gauged WZW model, and the deformation is totally encoded in the matrix $P$.

In \cite{AsyLambda}, the authors have studied the asymmetric $\lambda$-deformations of coset $SL(2,R)/U(1)$ and showed explicitly that the construction leads to a new integrable model. In this paper, we will apply these procedures to the cosets $SO(n+1)/SO(n)$ for $ n=2,3,\dots$.

\section{Asymmetric $\lambda$--deformed $SO(n+1)/SO(n)$}
\label{Main}
\renewcommand{\theequation}{3.\arabic{equation}}
\setcounter{equation}{0}

 The $\lambda$--deformation of the cosets  $SO(n+1)/SO(n)$ with $\lambda$-deformations for $n=2,3,4,5$  in the vector gauge cases have been constructed in \cite{MoreLmb,DSTLmbAdS5}. These corresponding deformed geometries can be promoted to integrable backgrounds of string theory and their dynamical properties are also analyzed in \cite{Lunin:2018vsn}.

 In the following discussion we choose the generators of $SO(n+1)$ to be
 \be\label{Generators} T_{ij}=\frac{i}{\sqrt{2}}(E_{ij}-E_{ji}) \ee
 and embed the subgroup $SO(n)$ as
 \be\label{SubgroupEmbed}
 t_A&=(\underbrace{T_{n,n+1};T_{n-1,n},T_{n-1,n+1};...,T_{23},...,T_{2,n+1}}_a;\underbrace{T_{12},...,T_{1,n+1}}_{\alpha})\\
 &\equiv(t_a,t_\alpha)
 \ee
 Before performing the deformation procedures we need to solve the outer automorphism group first. Besides that the other preparatory work is to find the coset representative for $SO(n+1)/SO(n)$ in the axial gauge.

\subsection{The outer automorphism group}
The outer automorphism group of a simply connected group corresponds to the symmetry of its Dynkin diagram. For simple Lie algebras, only the Dynkin diagrams of types $A_n$, $D_n$, and $E_6$ admit non-trivial symmetries. Therefore for orthogonal groups only the groups $SO(2n)$ with  $n\geq 2$ have non-trivial outer automorphism groups. The forms of the transformation $W$ are computed in Appendix \ref{AppendixW}. Up to inner automorphism the final results  are 
\be\label{Outer}
W(T_{ij})=
\left\{
\begin{array}{l}
	-T_{ij}, \quad j= 2 n \\
	T_{ij}\ \ \ , \quad j\leq 2 n-1\\
\end{array}
\right.
,\quad i<j,\text{ and } n\geq 2
\ee
Notice that $W$ is the diagonal matrix and $W=W^T=W^{-1}$.  This choice of gauge can be viewed as a higher dimensional generalization of the axial gauge used in \cite{Fradkin:1991ie}. 

\par

\subsection{The coset representative}
An element $g_{n+1} \in SO(n+1)$ can be decomposed as \cite{Bars:1991zt}
\be
g_{n+1}=H_nt_n
\ee
where
\bea
H_n=\left[
\begin{array}{cc}
	1 & 0 \\
	0 & h \\
\end{array}
\right],\quad t=\left[
\begin{array}{cc}
	b-1 & bV^T \\
	-bV & I_{n\times n}-bVV^T \\
\end{array}
\right],
\eea
with 
\bea
h\in SO(n)
, \quad b=\frac{2}{1+V^T V},
\eea
and $V$ is a $n$-vector. 
The $SO(n)$ gauge rotation $R_n$ acts as \eqref{GaugeTrans} 
\bea 
&& g_{n+1}'= R_n g_{n+1}\tilde{R}_n^{-1}=R_n H_n \tilde{R}_n^{-1}\tilde{R}_n t_n \tilde{R}_n^{-1}\equiv H_n't_n',\quad  V'_n=\tilde{R}_n V_n.
\eea 
We pick a convenient gauge such that  $V'_{n}=(v_{n-1},0,\dots,0)$ such that $t'_n$ is invariant under the $SO(n-1)$ gauge rotation $R_{n-1}$ and $H_n' \equiv g_n \in SO(n)$. For $g_n$, a similar decomposition and gauge choice lead to $g_{n-1}$ and $t_{n-1}'$. Eventually we can fix all the gauge freedoms and end up with the coset representative
\bea\label{CosetRer}
g'_{n+1}=R_1 R_2... R_{n} g_{n+1} \tilde{R}_{n}^{-1}...\tilde{R}_2^{-1}\tilde{R}_1^{-1}=t'_1 t'_2 ... t'_{n}
\eea
where
\bea\label{Coor}
t'_{i+1}&=&\left[
\begin{array}{cc}
	b_{i+1}-1 & b_{i+1}V_{i+1}^T \\
	-b_{i+1}V_{i+1} & 1-b_{i+1}V_{i+1}V_{i+1}^T \\
\end{array}
\right]
=\left[\begin{array}{cccc}
	I_{(n-1-i) \times (n-1-i)} &  &  & \\
	&\frac{1-v_i^2}{1+v_i^2}  & \frac{2 v_i}{1+v_i^2} & \\
	& -\frac{-2 v_i}{1+v_i^2} & \frac{1-v_i^2}{1+v_i^2}  & \\
	&  &  &I_{i \times i} \\
\end{array}
\right]\nn
&=&\left[\begin{array}{cccc}
	I_{(n-1-i) \times (n-1-i)} &  &  & \\
	&\cos\theta_i  & -\sin\theta_i & \\
	& \sin\theta_i & \cos\theta_i  & \\
	&  &  &I_{i \times i}\\
\end{array}
\right]=e^{\sqrt{2}i\theta_{i}T_{n-i,n+1-i}}, ~\theta_i=[0,2\pi).
\eea
The advantage of this coset representative is that it does not depend on the outer automorphsim transformation $W$. Another possible coset representative can be found in \cite{Grigoriev:2007bu}.

\subsection{An Example: $SO(4)/SO(3)$}
Before the generic discussion, let us study the simplest example in details. This example exhibits all the essential features of the general situations. The metrics of the corresponding gauged WZW model ($\lambda=0$) on this coset have been found both with vector and axial gauge in\cite{Fradkin:1991ie}. For the gauged WZW model, it turns out the metrics with different gauges are connected via a coordinate transformation. We will show for $\lambda$--deformed model, under a coordinate transformation and an additional transformation of the parameter $\lambda$ the two metrics with different gauges are related  to each other.
According to the general discussion \eqref{Outer}, the non-trivial automorphism $W$ is 
\bea 
W:\quad (T_{14},T_{24},T_{34})\rightarrow -(T_{14},T_{24},T_{34}).
\eea 
In order to compare with the results in \cite{Fradkin:1991ie}, we parameterize the group $SO(4)$ as
\bea\label{SO4Rep}
g=ht_3',\quad h=e^{i(\tau-\theta)T_{34}}e^{2i\phi T_{23}}e^{i(\tau+\theta)T_{34}}.
\eea 

\subsubsection*{Vector gauge}
In this gauge, the coset representative is given by setting $\theta=0$ in \eqref{SO4Rep}. Substituting the coset representative into \eqref{Metric} leads the metric
\bea 
&&\label{VectorMetric}
ds^2_V=\frac{2\pi}{k}\left[\frac{1+\lambda^2}{1-\lambda^2}e_\alpha e^\alpha+\frac{2\lambda}{1-\lambda^2}e_\alpha J_V^{\alpha\beta}e_{\beta}\right],\quad e_1=\frac{dV_1}{1+V_1^2},\\
&&e_2=-\frac{d\phi \cos\tau+d\tau \cot\phi \sin\tau}{V_1},\quad e_3=-V_1(d\phi \sin\tau-d\tau \cot\phi \cos\tau)\nonumber.
\eea 
In the two forms $e_2$ and $e_3$ one can recognize that
\bea 
E_1=-d\phi \cos\tau-d\tau \cot\phi \sin\tau,\quad E_2=-d\phi \sin\tau+d\tau \cot\phi \cos\tau.
\eea 
are the two frames  of the vector gauged  $SO(3)/SO(2)$ model \cite{Fradkin:1991ie}. It is more convenient to introduce the new variables
\bea \label{NewVarVector}
x=\cos\tau \cos\phi,\quad y=\sin\tau \cos\phi.
\eea 
With these new variables both $e_i$ and $J^{\alpha\beta}$ have relatively simpler expressions:
\bea \label{VectorJ}
&&E_1=-\frac{dx}{\sqrt{1-x^2-y^2}},\quad E_2=-\frac{dy}{\sqrt{1-x^2-y^2}},\\
&&J_V=\left(
\begin{array}{ccc}
	2 x^2+2 y^2-1 & 2 x \sqrt{-x^2-y^2+1} & -2 y \sqrt{-x^2-y^2+1} \\
	2 x \sqrt{-x^2-y^2+1} & 1-2 x^2 & 2 x y \\
	-2 y \sqrt{-x^2-y^2+1} & 2 x y & 1-2 y^2 \\
\end{array}
\right).
\eea 

\subsubsection*{Axial gauge}
In this gauge, the coset representative is given by setting $\tau=0$ in \eqref{SO4Rep}. Substituting the coset representative into \eqref{Metric} leads the metric
\bea \label{AxialMetric}
&&ds^2_A=\frac{2\pi}{k}\left[\frac{1+\lambda^2}{1-\lambda^2}e_\alpha e^\alpha+\frac{2\lambda}{1-\lambda^2}e_\alpha J_A^{\alpha\beta}e_{\beta}\right],\quad e_1=\frac{dV_1}{1+V_1^2},\\
&& e_2=\frac{-d\phi \cos\theta+d\theta \tan\phi \sin\theta}{V_1},\quad e_3=-V_1(d\phi \sin\theta+d\theta\tan\phi \cos\theta).
\eea 
In $e_2$ and $e_3$ again one can recognize that
\bea 
E_1=\text{d$\theta$} \sin (\theta) \tan (\phi )-\text{d$\phi $} \cos (\theta ),\quad E_2=-\text{d$\theta $} \cos (\theta ) \tan (\phi )-\text{d$\phi $} \sin (\theta )
\eea 
are frames of the axial gauged  $SO(3)/SO(2)$ model \cite{Fradkin:1991ie}. Similarly introducing the convenient variables 
\bea 
x=\cos\theta \sin\phi,\quad y=\sin\theta \sin\phi,
\eea 
we have
\bea 
&&E_1=-\frac{dx}{\sqrt{1-x^2-y^2}},\quad E_2=-\frac{dy}{\sqrt{1-x^2-y^2}},\\
&&J_A=-\left(
\begin{array}{ccc}
	2 x^2+2 y^2-1 & 2 x \sqrt{-x^2-y^2+1} & -2 y \sqrt{-x^2-y^2+1} \\
	2 x \sqrt{-x^2-y^2+1} & 1-2 x^2 & 2 x y \\
	-2 y \sqrt{-x^2-y^2+1} & 2 x y & 1-2 y^2 \\
\end{array}
\right).
\eea 
Comparing with results \eqref{VectorMetric} \eqref{VectorJ}, we find when $\lambda=0$ these two metrics are equivalent up to a change of coordinates:
\bea \label{ChangeCor}
\tau\rightarrow \theta,\quad \phi\rightarrow \frac{\pi}{2}-\phi.
\eea 
The equivalence of vector gauged and axial gauged model is called``self-dual" in \cite{Fradkin:1991ie}.
When $\lambda\neq0$, the metric $ds^2_A$ can be obtained from  $ds^2_V$ by performing \eqref{ChangeCor} and a $Z_2$ transformation on the deformation parameter:
\bea \label{Z2}
\lambda \rightarrow -\lambda.
\eea 
This additional $Z_2$ transformation is non-trivial for this model. One way to show this is to solve the spectrum of scalar on the deformed geometry following \cite{Lunin:2018vsn}. We will give a simple example in the next section.  In the next section, we will provide this kind of $Z_2$ transformation which transforms the deformed geometries in vector gauge into the ones in axial gauge for all the deformed cosets $SO(n+1)/SO(n)$.

\subsubsection*{Scalar field}
\label{Spectrum}

Even though the geometry corresponding to $\lambda$--deformed $SO(4)/SO(3)$ coset has no isometries, an algebraic method based on group theory is proposed in \cite{Lunin:2018vsn}.  In this appendix, we use their method to give a simple example showing that  the $Z_2$ transformation on the parameter $\lambda$ is physical. According to \cite{Lunin:2018vsn}, the scalar field equation on the deformed geometry reduces to a second order differential equation known as the Heun's equation for polynomial ansatz $Q(z)$:
\bea \label{Heun}
Q''(z)+\left[\frac{1}{z}+\frac{1+2L_2}{z-1}-\frac{1+2L_1}{z-c}\right]Q'(z)+\frac{(L_1-L_2)^2(z-h)}{z(z-1)(z-c)}Q(z)=0,
\eea 
where the parameters $(c,h)$ of the Heun's equation are given by
\bea 
c=\frac{1+\kappa}{2\kappa},~\kappa=\frac{2\lambda}{1+\lambda^2},~~h=\frac{(1-2c)L_1+(1-c)L_1^2+cL_2}{(L_1-L_2)^2}+\frac{\Lambda}{8\kappa(L_1-L_2)^2},
\eea 
and $(L_1,L_2)$ are integers labeling the irreducible representation of $SO(4)$. The spectrum $\Lambda$ is determined from the truncation condition of the polynomial ansatz solution of the Heun's equation. Assume the polynomial ansatz truncates at power $p$, the truncation conditions are
\bea
p=L_1-L_2
\eea 
and
\bea\label{detS3}
\mbox{det}(M_{p+1})=0,\ \mbox{where}\  M_{p+1}=\begin{bmatrix}
	A_0&-1&0&0&0&\dots &\dots &\dots\\
	B_1&A_1&-1&0&0&\dots &\dots &\dots\\
	0&B_2&A_2&-1&0&\dots &\dots &\dots\\
	\vdots & \vdots & \vdots & \vdots & \vdots &\dots &\vdots &\vdots &\\
	\dots & \dots &\dots &\dots &\dots & \dots & A_{p-1}&-1\\
	0&	0&	0&	0&	0&\dots & B_p& A_p
\end{bmatrix}\,.
\eea
where the matrix elements are defined by
\bea 
&&A_n=\frac{(1+c)n^2+[(1+2L_2)c-(1+2L_1)]n}{c(n+1)^2}+
\frac{(L_1-L_2)^2 h}{c(n+1)^2},\\
&&B_n=-\frac{(n-1+L_2-L_1)^2}{c(n+1)^2}\nonumber.
\eea 
Considering the example with $(L_2,L_1)=(2,1)$, \eqref{detS3} leads to
\bea 
&&8 (3 \kappa-5) \Lambda +48 (\kappa-7) (\kappa-1)+\Lambda ^2=0,\nn
&&\Lambda_{\pm}=4(5-3\kappa\pm\sqrt{4+6(\kappa-1)\kappa}),
\eea 
which are not equivalent under the transformation $\lambda\rightarrow -\lambda$.

\subsection{The $Z_2$ transformation}
In the section, we will not restrict ourselves to $SO(2n)/SO(2n-1)$ but consider the general coset $SO(n+1)/SO(n)$ model with the automorphism transformation
\bea\label{Auto}
W(T_{ij})=
\left\{
\begin{array}{l}
	-T_{ij}, \quad j= n+1 \\
	T_{ij}\ \ \ , \quad j\leq n\\
\end{array}
\right.
,\quad i<j,\text{ and } n\geq 2,
\eea
even though for the case when $n$ is even the transformation is just a gauge transformation. The deformed geometries \eqref{Metric} depend on $W$ in a rather complicate way. However 
we prove the deformed geometries can be transformed back to the symmetric $\lambda$--deformed ($W=I$) ones by the $Z_2$ transformation\footnote{up to some other coordinate transformations.}:
\bea \label{Z2Trans}
  \theta_i\rightarrow\pi-\theta_i,\quad \lambda\rightarrow-\lambda,
\eea 
where the coordinates $\theta_i$ of the target space are defined in \eqref{Coor}. The proofs are presented in the Appendix \ref{proof}. One prediction of our results is that the spectrum of scalar field on the $\lambda$--deformed $SO(2n+1)/SO(2n)$  is invariant under $\lambda\rightarrow -\lambda$ since $W$ is a gauge transformation in this case.

\subsection{Recursion relations}
Recall the deformed geometries are given by \eqref{FinalMetric}:
\bea 
&&ds^2=\frac{k}{2\pi}\frac{1-\lambda^2}{\lambda^2}e^T P^TP e,~~e_\alpha=L_\alpha-D^T_{a\alpha }(D_{ab}-W_{ab})^{-T}L_b\nn
&& (P^{-T})_{\alpha\beta}=\left[D_{\alpha\beta}-D_{\alpha b}(D_{ab}-W_{ab})^{-1}D_{a\beta}-\frac{1}{\lambda}W_{\alpha\beta}\right].
\eea
 In the proof the $Z_2$ transformation, we have obtained recursion relations for $e_\alpha$ ,  $D_{ab}$ and $P$. However, these relations are involved with cumbersome inversions of matrices. In this section, we will provide new recursion relations which only consist of matrix additions and multiplications. By the $Z_2$ transformation, we can set $W=I$ and rewrite the metric in a similar form as \eqref{VectorMetric}:
\be
ds^2&=\frac{k}{2\pi}\frac{1-\lambda^2}{\lambda^2} e^T P^T P e\equiv\frac{k}{2\pi}\(\frac{1+\lambda^2}{1-\lambda^2} e^T e+\frac{2\lambda}{1-\lambda^2} e^T J e\),
\ee
where 
\be\label{Jmatrix}
J=\frac{1-\lambda^2}{2\lambda }\left(\frac{1-\lambda^2}{\lambda^2}P^T P - \frac{1+\lambda^2}{1-\lambda^2} I\right).
\ee
Introducing a new quantity 
\bea 
Q\equiv d_4-d_3(d_1-1)^{-1}d_2,
\eea
where $d_i$ are block elements in the matrix $D_{AB}$
\be
D\equiv\left[
\begin{array}{cc}
	d_1 & d_2 \\
	d_3 & d_4 \\
\end{array} \right]. 
\ee
The quantity $Q$ satisfies $Q^TQ=QQ^T=I$ \cite{DSTLmbAdS5}. In Appendix \ref{AppendixQ} we show  $Q=Q^T$ and $Q=J$.  Here we present the new recursion relations for $e_\alpha$ and $Q$ .
The recursion relation of Q is given by\footnote{The subscript of $X\s{i+1}$ denotes that the quantity $X\s{i+1}$ is defined for the coset $SO(i+1)/SO(i)$.  } 
\be\label{Qre}
Q\s{n+1}
&=\left[
\begin{array}{cc}
	\cos\ta_{n-2}&
	\left[
	\begin{array}{c}
		\sin\ta_{n-2}\cos\ta_{n-3} \\
		\sin\ta_{n-2}\sin\ta_{n-3}\cos\ta_{n-4}\\
		\vdots\\
		\sin\ta_{n-2}\sin\ta_{n-3}\ldots \sin\ta_{1}\cos\ta_{0}\\
		\sin\ta_{n-2}\sin\ta_{n-3}\ldots \sin\ta_{1}\sin\ta_{0}\\
	\end{array}
	\right]^T\\
	\left[
	\begin{array}{c}
		\sin\ta_{n-2}\cos\ta_{n-3} \\
		\sin\ta_{n-2}\sin\ta_{n-3}\cos\ta_{n-4}\\
		\vdots\\
		\sin\ta_{n-2}\sin\ta_{n-3}\ldots \sin\ta_{1}\cos\ta_{0}\\
		\sin\ta_{n-2}\sin\ta_{n-3}\ldots \sin\ta_{1}\sin\ta_{0}\\
	\end{array}
	\right]
	& Q\s n
	\left[
	\begin{array}{cc}
		-\cos\ta_{n-2} &  \\
		& {Q\s {n-1}} \\
	\end{array}
	\right]
	Q\s n\\
\end{array}
\right]\\
\ee
with the first two cases
\be
&Q\s{3}=\left[
\begin{array}{cc}
	\cos \ta_0 & \sin \ta_0 \\
	\sin \ta_0 & -\cos \ta_0 \\
\end{array}
\right],\\
&Q\s{4}=\left[
\begin{array}{ccc}
	\cos \ta_1 & \cos \ta_0 \sin \ta_1 & \sin \ta_0 \sin \ta_1 \\
	\cos \ta_0 \sin \ta_1 & \sin ^2\ta_0-\cos ^2\ta_0 \cos \ta_1 & -2 \cos \ta_0 \cos ^2\left(\frac{\ta_1}{2}\right) \sin \ta_0 \\
	\sin \ta_0 \sin \ta_1 & -2 \cos \ta_0 \cos ^2\left(\frac{\ta_1}{2}\right) \sin \ta_0 & \cos ^2\ta_0-\cos \ta_1 \sin ^2\ta_0 \\
\end{array}
\right].\\
\ee

 For the frames $e_\alpha$, since the expression of $L$ has been shown in \eqref{FrameExp}, we only need to consider $(d_1\s{n+1}-1)^{-1}d_2\s{n+1}$.  The block matrix  inversion formula gives
 \be
(d_1\s{n+1}-1)^{-1}&=\left[
\begin{array}{cc}
	\ast & -(d_1\s{n}-1)^{-1}d_2\s{n}\cos\ta_{n-1}\(\cos\ta_{n-1} Q\s n -1\)^{-1} \\
	\ast & \(\cos\ta_{n-1} Q\s n -1\)^{-1} \\
\end{array}
\right],\\
\ee
where $\ast$ stands for the irrelevant elements.  Substituting the identity \eqref{DnBlock} we get
\be\label{eq5.5}
&(d_1\s{n+1}-1)^{-1}d_2\s{n+1}\\
=&\tan\ta_{n-1}\[1+(d_1\s{n+1}-1)^{-1}\]C_{n-1}\\
=&\tan\ta_{n-1}\left[
\begin{array}{cc}
	0 & -(d_1\s{n}-1)^{-1}d_2\s{n}\cos\ta_{n-1}\(\cos\ta_{n-1} Q\s n -1\)^{-1} \\
	0 & \(\cos\ta_{n-1} Q\s n -1\)^{-1}+1 \\
\end{array}
\right]\\
=&\[
\begin{array}{cc}
0 & (d_1\s{n}-1)^{-1}d_2\s{n}\frac{1+Q\s n \cos\ta_{n-1}}{\sin\ta_{n-1}} \\
0 & -\frac{\cos\ta_{n-1}+Q\s n }{\sin\ta_{n-1}} \\
\end{array}\], 
\ee
with the first case
\bea 
(d_1\s{3}-1)^{-1}d_2\s{3}=\left[ 0 , -\cot \left({\ta_1}/{2}\right) \right].
\eea 
In the end let us derive the recursion relation of the dilaton \eqref{dilaton1}($W=I$):
\bea 
e^{-2\Phi}=e^{-2\Phi_0}\mbox{det}(d_1-1).
\eea 
Substituting the recursion relation of $d_1$ \eqref{Drelation} leads to
\bea 
\mbox{det}(d_1\s{n+1}-1)=\mbox{det}(d_1\s{n}-1)\mbox{det}[1-\cos\theta_{n-1}Q 
\s n].
\eea 
To evaluate the second determinant we first observe that the eigenvalues of $Q$ can only be $\pm 1$ due to $QQ^T=I$. Then taking trace on both sides of the recursion relation \eqref{Qre}, one finds
\bea 
\mbox{Tr}[Q\s{n+1}]=\mbox{Tr}[Q\s{n-1}],  \mbox{ or } \mbox{Tr}[Q\s{2n-1}]=0,~\mbox{Tr}[Q\s{2n}]=1,~n=2,3\dots.
\eea 
Combining the  two facts we get
\bea 
&&\mbox{det}[1-\cos\theta_{2k-1}Q\s {2k}]=(1-\cos\theta_{2k-1})\sin^{2k}\theta_{2k-1},\nn &&\mbox{det}[1-\cos\theta_{2k-2}Q\s {2k-1}]=\sin^{2(k-1)}\theta_{2k-2},~k=2,3,\dots,
\eea 
which complete our recursion relations.
\section{Summary}
In this article we explored the asymmetric $\lambda$--deformation introduced in \cite{AsyLambda}. For the $SO(n+1)/SO(n)$ coset model we found a $Z_2$ transformation \eqref{Z2Trans} which maps the asymmetric $\lambda$--deformation to the symmetric $\lambda$--deformations . When the deformation parameter $\lambda$ vanishes, these two different deformed models reduce to the gauged WZW models in the axial and vector gauge, respectively. The gauged WZW models in different gauges are dual to each other.  We gave evidences to show that the deformation break this duality so that  the asymmetric $\lambda$--deformation  leads to new integrable models. Furthermore, we construct the resulting geometries for arbitrary $n$ recursively \eqref{Qre}, \eqref{eq5.5} and \eqref{FrameExp}. It would be interesting to extend the current results to the cases of supergroups so that the integrable deformation of string theory on $AdS_p\times S^p$ can be constructed and studied.


\section*{Acknowledgments}
The work was in part supported by NSFC Grant  No.~11335012, No.~11325522 and No. 11735001.

\appendix
\section{The outer automorphism of $so(n)$}
\label{AppendixW}
\renewcommand{\theequation}{A.\arabic{equation}}
\setcounter{equation}{0}
For the Lie algebra, $W$ is called the outer automorphism transformation if there is no $x^A\in \mathcal{R}$ such that $W=exp\(x^A ad_{t_A}\)$, where $ad$ means the adjoint representation.
Let us consider $so(2n), n\geq3$ in the Cartan-Weyl basis. Its simple roots are
\be
&\alpha_i=e_i-e_{i+1},\quad i=1,2,...,n-1\\
&\alpha_n=e_{n-1}+e_{n}\\
\ee
and the generators are
\be
&H_i=-\sqrt{2}T_{(2j-1)(2j)},\\
&E_{\alpha_i}=-\frac{1}{\sqrt{2}}\[T_{(2i-1)(2i+1)}+T_{(2i)(2i+2)}+i T_{(2i)(2i+1)}-i T_{(2i-1)(2i+2)}\],\quad i=1,2,...,n-2,\\
&E_{\alpha_{n-1}}=-\frac{1}{\sqrt{2}}\[T_{(2n-3)(2n-1)}+T_{(2n-2)(2n)}+i T_{(2n-2)(2n-1)}-i T_{(2n-3)(2n)}\],\\
&E_{\alpha_{n}}=-\frac{1}{\sqrt{2}}\[T_{(2n-3)(2n-1)}-T_{(2n-2)(2n)}+i T_{(2n-2)(2n-1)}+i T_{(2n-3)(2n)}\],
\ee
where $T_{ij}$ are denoted by \eqref{Generators}. According to the symmetry of the Dynkin diagrams of $D_n$, the outer automorphism $W$ are \cite{Ganor:2002ya}
\be
&W(E_{\alpha_i})=W(E_{\alpha_i}),\quad i=1,2,...,n-2,\\
&W(E_{\alpha_{n-1}})=W(E_{\alpha_n}),\quad W(E_{\alpha_n})=W(E_{\alpha_{n-1}}). 
\ee
Notice that $W$ is a linear transformation. Therefore,
\be\label{eqW1}
&W(T_{(2i-1)(2i+1)})=T_{(2i-1)(2i+1)},\quad W(T_{(2i)(2i+1)})=T_{(2i)(2i+1)},\quad i=1,2,...,n-1\\
&W(T_{(2i)(2i+2)})=T_{(2i)(2i+2)},\quad W(T_{(2i-1)(2i+2)})=T_{(2i-1)(2i+2)},\quad i=1,2,...,n-2\\
&W(T_{(2n-2)(2n)})=-T_{(2n-2)(2n)},\quad W(T_{(2n-3)(2n)})=-T_{(2n-3)(2n)}.\\
\ee
By \eqref{eqW1} and commutators of $SO(n)$ group
\be\label{eqcomm}
\[T_{AB},T_{CD}\]=\frac{i}{\sqrt{2}}\(\delta_{BC}T_{AD}+\delta_{AD}T_{BC}-\delta_{AC}T_{BD}-\delta_{BD}T_{AC}\).
\ee
we can get an outer automorphism transformation,
\be\label{eqW2}
W(T_{ij})=
\left\{
\begin{array}{l}
	-T_{ij}, \quad j= 2 n \\
	T_{ij}\ \ \ , \quad j\leq 2 n-1\\
\end{array}
\right.
,\quad i<j,\text{ and } n\geq 3. 
\ee\par
As for $so(4)$, which is isomorphism to $so(3)\oplus so(3)$. The generators are
\be
J_{+1}=\frac{1}{\sqrt{2}}(T_{23}+T_{14}),\quad J_{+1}=\frac{1}{\sqrt{2}}(T_{13}-T_{24}),\quad J_{+1}=\frac{1}{\sqrt{2}}(T_{12}+T_{34}),\\
J_{-1}=\frac{1}{\sqrt{2}}(T_{23}-T_{14}),\quad J_{-1}=\frac{1}{\sqrt{2}}(T_{13}+T_{24}),\quad J_{-1}=\frac{1}{\sqrt{2}}(T_{12}-T_{34}).\\
\ee
The commutators are
\be
&\[J_{+i},J_{+j}\]=i \epsilon_{ij}{}^{k}J_{+k},\quad \[J_{-i},J_{-j}\]=i \epsilon_{ij}{}^{k}J_{-k}, \quad 
\[J_{+i},J_{-j}\]=0. 
\ee
The adjoint representations of the generators are
\be
&J_{+i}=\left[
\begin{array}{cc}
	\sigma_i &0 \\
	0 &0 \\
\end{array}
\right]
=\sigma_i \oplus 0, \hspace{3ex}
J_{-i}=\left[
\begin{array}{cc}
	0 &0 \\
	0 &\sigma_i \\
\end{array}
\right]
=0 \oplus \sigma_i, \\
\ee
where $\sigma_i$ are Pauli matrices in the adjoint representation, that is,
\be
\sigma_1=\left[
\begin{array}{ccc}
	&  &  \\
	&  & i \\
	& -i &  \\
\end{array}
\right],\quad
\sigma_2=\left[
\begin{array}{ccc}
	&  & -i \\
	&  &  \\
	i  &  &  \\
\end{array}
\right],\quad
\sigma_3=\left[
\begin{array}{ccc}
	&i  &  \\
	-i  &  &  \\
	&  &  \\
\end{array}
\right]. \quad
\ee
Consider an automorphism $W$,
\be
W(J_{+i})=J_{-i},\quad W(J_{-i})=J_{+i}. 
\ee
Under the basis $J=(J_+, J_-)$,
\be
W=\left(
\begin{array}{cc}
	0 & I \\
	I & 0 \\
\end{array}
\right). 
\ee
However, by the definition of outer automorphism,
\be
e^{x^A ad_{J_A}}=e^{(x_+^i \sigma_i)\oplus (x_-^j\sigma_j)}=e^{(x_+^i \sigma_i)}\oplus e^{(x_-^j\sigma_j)}=
\left(
\begin{array}{cc}
	e^{x_+^i \sigma_i} & 0 \\
	0 & e^{x_-^i \sigma_i} \\
\end{array}
\right), 
\ee
 there is no solution about $W=e^{x^A ad_{J_A}}$,  so $W$ is an outer automorphism. Under the basis $T_{ij}$,
\be\label{eqW3}
W(T_{ij})=
\left\{
\begin{array}{l}
	-T_{ij}, \quad j= 4 \\
	T_{ij}\ \ \ , \quad j\leq 3\\
\end{array}
\right.
,\quad i<j. 
\ee

\section{Proof of the $Z_2$ transformation}
\label{proof}
\renewcommand{\theequation}{B.\arabic{equation}}
\setcounter{equation}{0}
In this appendix we will prove that under the combination $W$ and \eqref{Z2Trans} the geometry \eqref{Metric} is invariant. Our main method is the mathematical induction. We begin with proving some import properties of matrix  $D_{AB}$ defined in \eqref{MCforms} under the coset representative \eqref{CosetRer}:

\bea
 g=t_1't_2'\dots t_n'
\eea
The matrix $D$ can be decomposed as
\bea
D_{AB}=D^1_{AC_1}D^2_{C_1C_2}...D^n_{C_{n-1}B}, 
\eea
where 
\bea
D^i_{AB} \equiv \mbox{Tr} [T_{A}t'_i T_{B}{t'_i}^{-1}]\equiv
\begin{blockarray}{ccc}
a & \alpha & \\
\begin{block}{[cc]c}
d^i_1 & d^i_2 & a \\
d^i_3 & d^i_4 & \alpha \\
\end{block}
\end{blockarray} 
\eea
with $t'_i=e^{\sqrt{2}i\theta_{i-1}T_{n+1-i,n+2-i}}$.
Because the generators $T_A$ normalized as \eqref{Generators} satisfy the identities
\bea\label{GenIdentity}
\left\{
\begin{array}{l}
	e^{-\sqrt{2}i T_{i,i+1}\ta}T_{i+1,j}e^{\sqrt{2}i T_{i,i+1}\ta}=\cos\ta T_{i+1,j}+\sin\ta T_{i,j}, \\
	e^{-\sqrt{2}i T_{i,i+1}\ta}T_{i,j}e^{\sqrt{2}i T_{i,i+1}\ta}=\cos\ta T_{i,j}-\sin\ta T_{i+1,j}, \\
	e^{-\sqrt{2}i T_{i,i+1}\ta}T_{m,n}e^{\sqrt{2}i T_{i,i+1}\ta}=T_{m,n},\quad \mbox{for other } T_{m,n}.\\
\end{array}
\right.
\eea
All the matrices $D^i$ are block-diagonal (i.e., $d^i_2=d^i_3=0$) except for  $D^n$ which has the form
\be\label{DnBlock}
D^n=
\begin{blockarray}{cccc}
	\overbrace{\hskip 2.5cm}^{a'} &\overbrace{\hskip 2.5cm}^{ T_{23},...,T_{2(n+1)}} & \overbrace{\hskip 1.5cm}^{T_{12}} & \overbrace{\hskip 2.5cm}^{T_{13},..,T_{1(n+1)}} \\
	\begin{block}{[cc|cc]}
		I_{(n-1)(n-2)/2} & 0 & 0 & 0  \\
		0 & cos\theta_{n-1}I_{n-1} & 0 & sin\theta_{n-1}I_{n-1} \\
		\BAhhline{----}
		0 & 0 & 1 & 0  \\
		0 & -sin\theta_{n-1}I_{n-1} & 0 & cos\theta_{n-1}I_{n-1}  \\
	\end{block}
	\BAmulticolumn{2}{c}{\underbrace{\hskip 5cm}_{a=n(n-1)/2 }}&\BAmulticolumn{2}{c}{\underbrace{\hskip 4cm}_{ \alpha=n} }\\
\end{blockarray}.
\ee
Therefore, the matrix $D$ can be written as
\be
D&=D^1D^2...D^{n-1}D^n\\
&=\left[
\begin{array}{cc}
	d^1_1 & 0 \\
	0 & d^1_4 \\
\end{array}
\right]
\left[
\begin{array}{cc}
	d^2_1 & 0 \\
	0 & d^2_4 \\
\end{array}
\right]
...
\left[
\begin{array}{cc}
	d^{n-1}_1 & 0 \\
	0 & d^{n-1}_4 \\
\end{array}
\right]
\left[
\begin{array}{cc}
	d^n_1 & d^n_2 \\
	d^n_3 & d^n_4 \\
\end{array}
\right]\\
&=\left[
\begin{array}{cc}
	d^1_1...d^{n-1}_1d^n_1 & d^1_1...d^{n-1}_1d^n_2 \\
	d^1_4...d^{n-1}_4d^n_3 & d^1_4...d^{n-1}_4d^n_4 \\
\end{array}
\right]\\
&\equiv
\left[
\begin{array}{cc}
	d_1 & d_2 \\
	d_3 & d_4 \\
\end{array}
\right],\quad d_2=d_1(d_1^n)^{-1}d_2^n,\quad d_3=d_4(d_4^n)^{-1}d_3^n .
\ee
Using the matrix \eqref{DnBlock}, we can expression off-diagonal blocks with the diagonal blocks as
\bea \label{DRe}
&&d_2=\tan\theta_{n-1}d_1 C_{n-1},\quad d_3=-\tan\theta_{n-1}d_4 C^T_{n-1},\nn
&&C_{n-1}=
\underbrace{\left[
	\begin{array}{cc}
		0 & 0 \\
		0 & I_{n-1} \\
	\end{array}
	\right]}_{ \alpha=n}
\Bigg\}{}_{a=n(n-1)/2}.
\eea 
For the diagonal blocks $d_1$ and $d_4$ we provide  recursion relationships:
\be\label{Recusion}
d_4|_{n+1}
&=d_4^1\s{n+1}...d_4^{n-1}\s{n+1}d_4^n\s{n+1}\\
&=\left[
\begin{array}{cc}
	I_{n-2} &  \\
	& R(\ta_0) \\
\end{array}
\right]
...
\left[
\begin{array}{cc}
	R(\ta_{n-2}) &  \\
	& I_{n-2} \\
\end{array}
\right]
\left[
\begin{array}{cc}
	1 &  \\
	& \cos\ta_{n-1}I_{n-1} \\
\end{array}
\right]\\
&=  \left[
\begin{array}{cc}
	1 &  \\
	& d_4\s  n \\
\end{array}
\right]
\left[
\begin{array}{cc}
	R(\ta_{n-2}) &  \\
	& \frac{I_{n-2}}{\cos\ta_{n-2}} \\
\end{array}
\right]
\left[
\begin{array}{cc}
	1 &  \\
	& \cos\ta_{n-1}I_{n-1} \\
\end{array}
\right],~~R(\theta)=\begin{bmatrix}
	\cos\theta & -\sin\theta\\
	\sin\theta & \cos\theta\end{bmatrix}
\ee
and
\be\label{Drelation}
d_1\s{n+1}
&=d_1^1\s{n+1}...d_1^{n-1}\s{n+1}d_1^n\s{n+1}=D^1\s{n}...D^{n-1}\s{n}d_1^n\s{n+1}\\
&=D\s{n}d_1^n\s{n+1}=\left[
\begin{array}{cc}
	d_1\s n & d_2\s n \cos\ta_{n-1} \\
	d_3\s n & d_4\s n \cos\ta_{n-1} \\
\end{array}
\right].
\ee

\vskip 0.5cm

\noindent {\bf The frames $e_\alpha$}

Let us study how the frames 
\bea 
e_\alpha=L_\alpha-d_2^T(d_1^T-W_a)^{-1}L_a
\eea 
transform under \eqref{Z2Trans}. Recalling the identities \eqref{GenIdentity}, the left Maurer-Cartan forms can be calculated explicitly and the final expression can be written in a compact way
\be\label{FrameExp}
L=\sqrt{2}\left[
\begin{array}{cccccc}
	\ast&d\ta_{n-1} &&...&...& d\ta_0  \sin\ta_{n-1}...\sin\ta_1\\
	\multicolumn{6}{c}{\vdots} \\
	&& \ast& d\ta_2 \cos\ta_3 & d\ta_1 \cos\ta_3 \sin\ta_2 & d\ta_0 \cos\ta_3 \sin\ta_2\sin\ta_1 \\
	&& & \ast& d\ta_1 \cos\ta_2 & d\ta_0 \cos\ta_2 \sin\ta_1 \\
	&& &  & \ast & d\ta_0 \cos\ta_1 \\
	&& &  &  & \ast \\
\end{array}
\right]. 
\ee
The expression means that the element of the matrix in the (i,j) position is the element of L, whose position in $L$ is the same as the position of $T_{i,j}$ in $t_{A}$. So  under the transformation \eqref{Z2Trans} the forms transform as
\bea \label{FrameTran}
L_\alpha \rightarrow -L_\alpha,\quad L_a\rightarrow L_a.
\eea 
Next we focus on the term $W_a d_1$ in the combination
\bea 
(d_1-W_a)^{-1}d_2=[1-(W_a d_1)^{-1}]^{-1}d_1^{-1}d_2=\tan\theta_{n-1}[1-(W_a d_1)^{-1}]^{-1} C_{n-1},
\eea 
in which we already have $\tan\theta_{n-1}\rightarrow -\tan\theta_{n-1}$ under $Z_2$.
Using the recursion relation \eqref{Recusion} and \eqref{Drelation} , one can get
\bea 
&&W_a d_1\s {n+1}
=\left[
\begin{array}{cc}
	W_{a'} d_1\s n & W_{a'} d_2\s n \cos\ta_{n-1} \\
	W_{\alpha'} d_3\s n & W_{\alpha'} d_4\s n \cos\ta_{n-1} \\
\end{array}
\right]\\
&&W_\alpha d_4\s {n+1}=\left[
\begin{array}{cc}
	1 &  \\
	& W_{\alpha'} d_4\s n cos\ta_{n-1}\\
\end{array}
\right]
\left[
\begin{array}{ccc}
	cos\ta_{n-2} &-sin\ta_{n-2}cos\ta_{n-1} & \\
	\frac{sin\ta_{n-2}}{cos\ta_{n-1}}&cos\ta_{n-2}&\\
	& & \frac{I_{n-2}}{cos\ta_{n-2}} \\
\end{array}
\right]\nn
&&\cos\ta_{n-1}d_2 \s n=d_1\s n C_{n-2}\tan\theta_{n-2}\cos\ta_{n-1},~~\cos\ta_{n-1}d_3 \s n=-\tan\theta_{n-2}\cos\ta_{n-1}d_4C^T_{n-2},\nonumber
\eea 
where $a$ is the index of $so(n)$, $a'$ is the index of $so(n-1)$ and $\alpha'$ is the index of $so(n)-so(n-1)$. Observing that the \eqref{Z2Trans} has the same form for a generic $n$, below we will use an inductive proof to show that $W_a d_1$ is invariant under \eqref{Z2Trans}. Assume that $	W_{a} d_1\s n$ and $W_{\alpha} d_4\s n cos\ta_{n-1}$ are invariant one can find that $W_{a} d_1\s {n+1}$ and $W_{\alpha} d_4\s {n+1}cos\ta_{n}$ are also invariant from their expressions. For $n=2$, the result is simply proved by a direct substitution. Therefore, combining the transformations \eqref{FrameTran} we conclude under the $Z_2$ transformation, the frames transform as
\bea 
e_\alpha\rightarrow -e_\alpha.
\eea 

\vskip 0.5cm

\noindent {\bf The matrix $PP^T$}

Recall the definition of the matrix $P$
\bea 
P^{-T}=d_4-d_3(d_1-W_a)^{-1}d_2-\frac{1}{\lambda}W_\alpha.
\eea 
Using the identities $DD^T=WW^T=I$, one can show
\bea\label{Pidentity} 
\(P^{-1}+\frac{W_\alpha}{\lambda}\)\(P^{-T}+\frac{W_\alpha^{T}}{\lambda}\)=\(P^{-T}+\frac{W_\alpha^{T}}{\lambda}\)\(P^{-1}+\frac{W_\alpha}{\lambda}\)=1.
\eea 
Therefore, we  solve
\be
(P^T P)^{-1}
&=\left[d_4^T-\frac{1}{\lambda}W_{\alpha}-d_2^T(d_1^T-W_a)^{-1}d_3^T\right]\left[d_4-\frac{1}{\lambda}W^T_{\alpha}-d_3(d_1-W_a^T)^{-1}d_2\right]\\
&=1+\frac{1}{\lambda^2}-\frac{1}{\lambda}\[W_\alpha\(P^{-T}+\frac{W_\alpha^{T}}{\lambda}\)+\(P^{-1}+\frac{W_\alpha}{\lambda}\)W_\alpha^T\].
\ee
Substituting the relation \eqref{Drelation}, the first term in the bracket can be rewritten as
\be
\frac{1}{\lambda}W_\alpha\(P^{-T}+\frac{W_\alpha^{T}}{\lambda}\)
=\[\frac{1}{\lambda \cos\ta_n}\]\[W_a d_4\s{n+1}\cos\ta_n\]\[1+\tan\ta_{n-1}C^T_{n}(d_1-W_a^T)^{-1}d_2\]. 
\ee
The three parts,$\[...\],\[...\],\[...\]$, as shown before are invariant, respectively under $Z_2$. We conclude that under the $Z_2$ transformation, the matrix $PP^T$ is invariant. 
\vskip 0.5cm

\noindent {\bf The dilaton}

The deformed dilaton is given by 
\bea 
e^{-2\Phi}=e^{-2\Phi_0}\mbox{det}(N)\mbox{det}(W),\quad N=D-W\Omega.
\eea 
The determinant $\mbox{det}(W)=\pm 1$ can be absorbed into $\Phi_0$ so we focus on $\mbox{det}(N)$. To compute this determinant we rewrite matrix $N$ as
\bea 
&&N=\begin{bmatrix}
	d_1-W_a&d_2\\
	d_3&d_4-\lambda^{-1}W_\alpha
\end{bmatrix}=\begin{bmatrix}
d_1-W_a&0\\
d_3&I
\end{bmatrix}\begin{bmatrix}
I&(d_1-W_a)^{-1}d_2\\
0&P^{-T}
\end{bmatrix}
\eea 
Then
\bea \label{dilaton}
\mbox{det}(N)&=&\mbox{det}(W_a)\mbox{det}(1-W_ad_1)\mbox{det}[P^{-T}].
\eea 
The identity \eqref{Pidentity} implies that the eigenvalues of $P^{-1}$ are $\pm1  -W_\alpha \lambda^{-1}$ so that $\mbox{det}[P^{-T}]$ is a constant and can be absorbed into $\Phi_0$. Therefore after absorbing all the constants we end up with
\bea\label{dilaton1}
e^{-2\Phi}=e^{-2\Phi_0}\mbox{det}(1-W_ad_1).
\eea 
Using the previous results in the appendix, we conclude that the dilaton does not change under $Z_2$.

To summarize, in this appendix by calculating the transformations of the frames $e_\alpha$ , the deformation matrix $PP^T$ and $\mbox{det}(N)$ we have proved that deformed geometry is invariant under \eqref{Z2Trans}.

\section{Property of matrix $Q$}
\label{AppendixQ}
\renewcommand{\theequation}{C.\arabic{equation}}
\setcounter{equation}{0}
In this appendix we first derive the recursion relation for $Q$, and then prove $Q=Q^T$ by induction.

Recall the definition of the matrix $Q$:
\bea 
Q\s{n+1}=d_4\s{n+1}-d_3\s{n+1}(d_1\s{n+1}-1)^{-1}d_2\s{n+1}.
\eea 
Substituting \eqref{DRe} into the expression gives
\bea \label{Qexp}
Q\s{n+1}&=&d_4\s{n+1}\[1+\tan^2\ta_{n-1} C^T_{n-1}C_{n-1}+\tan^2\ta_{n-1} C^T_{n-1}(d_1\s{n+1}-1)^{-1}C_{n-1}\],\nn
&=&d_4\s{n+1}\(1+\tan^2\ta_{n-1}\left[
\begin{array}{cc}
0& 0 \\
0 & \(\cos\ta_{n-1} Q\s n -1\)^{-1}+1 \\
\end{array}
\right]\).
\eea 
To prove $Q=Q^T$ inductively, we first assume $Q\s i=Q^T\s i$, $\forall i<n+1$, then $Q\s n Q\s n^T=Q\s n^2=I$  and the matrix inversion in \eqref{Qexp} can be computed explicitly as
\bea\label{Qexp2} 
&&\(\cos\ta_{n-1} Q\s n -1\)^{-1}=-\frac{1+\cos\ta_{n-1} Q\s n}{\sin^2\ta_{n-1}},~Q\s{n+1}=d_4\s{n+1}\left[
\begin{array}{cc}
	1& 0 \\
	0 & -\frac{ Q\s n}{\cos\ta_{n-1}} \\
\end{array}
\right].
\eea 
Applying the recursion relation \eqref{Recusion} of $d_4$ we can cast $Q\s {n+1}$ into
\bea 
&&Q\s {n+1}=\left[
\begin{array}{cc}
\cos\ta_{n-2}& [\sin\ta_{n-2},0,...,0]Q\s n \\
	d_4\s n \left[
	\begin{array}{c}
		\sin\ta_{n-2} \\
		0 \\
		\vdots \\
		0 \\
	\end{array}
	\right]
	& -\cos\ta_{n-2} d_4\s n K_n Q\s n
\end{array}
\right]\equiv\begin{bmatrix}
	q_1&q_2\\q_3&q_4
	\end{bmatrix},
 \nn
&& K_n=  \begin{bmatrix}
	1 & 0 \\
	0 & \frac{I_{n-2}}{\cos^2\ta_{n-2}} \\
\end{bmatrix}. 
\eea 
The equation \eqref{Qexp2} and $Q\s n=Q^T\s n $ imply $(Q\s {n})_{i1}=(Q\s {n})_{1i}=(d_4\s {n})_{i1}$ so that $q_2=q_3$. Consider the analogue expression of \eqref{Qexp2} for $Q\s{n}$:
\bea 
Q\s{n}=d_4\s{n}\left[
\begin{array}{cc}
	1& 0 \\
	0 & -\frac{ Q\s {n-1}}{\cos\ta_{n-2}} \\
\end{array}
\right]\equiv d_4\s{n} K_n',\quad K'_n={K'}_n^T.
\eea 
Multiplying $K'_n$ on both sides the equation leads to
\bea
 Q\s{n}K'_n=d_4K_n.
\eea
Therefore we conclude $q_4$ and $Q\s{n+1}$ are symmetric. The direct calculation shows
\be
&Q\s{3}=\left[
\begin{array}{cc}
	\cos \ta_0 & \sin \ta_0 \\
	\sin \ta_0 & -\cos \ta_0 \\
\end{array}
\right],\\
&Q\s{4}=\left[
\begin{array}{ccc}
	\cos \ta_1 & \cos \ta_0 \sin \ta_1 & \sin \ta_0 \sin \ta_1 \\
	\cos \ta_0 \sin \ta_1 & \sin ^2\ta_0-\cos ^2\ta_0 \cos \ta_1 & -2 \cos \ta_0 \cos ^2\left(\frac{\ta_1}{2}\right) \sin \ta_0 \\
	\sin \ta_0 \sin \ta_1 & -2 \cos \ta_0 \cos ^2\left(\frac{\ta_1}{2}\right) \sin \ta_0 & \cos ^2\ta_0-\cos \ta_1 \sin ^2\ta_0 \\
\end{array}
\right],\\
\ee
which are both symmetric then we finish our inductive proof.

From the definition of J \eqref{Jmatrix}, we can rewrite $J$ by $Q$,
\be\label{J}
J&=\(\xi-2 Q\)^{-1}\(\xi Q -2\),\quad \xi=\frac{1+\lambda^2}{\lambda}. \\
\ee
Taking the derivative with respect to $\xi$ leads to
\be
\frac{d J}{d \xi}=-2(\xi-2 Q)^{-2}\(Q^2-1\)=0.
\ee
Here we have used the identity $Q^2=1$. Therefore, $J$ does not depend on $\lambda$.\par
Furthermore multiplying $\(\xi-2 Q\)$ on both sides gives
\be
\xi\(J-Q\)-2\(J Q-1\)=0.
\ee
Because both $J$ and $Q$ do not depend on $\xi$, we conclude 
\be J=Q. \ee


\newpage
\begin{thebibliography}{30}


\bibitem{AsyLambda}
S.~Driezen, A.~Sevrin and D.~C.~Thompson,
``Integrable asymmetric $\lambda$-deformations,''
JHEP {\bf 1904}, 094 (2019)
[arXiv:1902.04142 [hep-th]].

\bibitem{BosonicYangBaxter}
C.~Klimcik,
``Yang-Baxter sigma models and dS/AdS T duality,''
JHEP {\bf 0212}, 051 (2002), hep-th/0210095; \\
%
C.~Klimcik,
``On integrability of the Yang-Baxter sigma-model,''
J.\ Math.\ Phys.\  {\bf 50}, 043508 (2009), arXiv:0802.3518;\\
%
C.~Klimcik,
``Integrability of the bi-Yang-Baxter sigma-model,''
Lett.\ Math.\ Phys.\  {\bf 104}, 1095 (2014), arXiv:1402.2105;\\
C.~Klimcik,
``Poisson--Lie T-duals of the bi-Yang--Baxter models,''
Phys.\ Lett.\ B {\bf 760}, 345 (2016),
arXiv:1606.03016 [hep-th].
\bibitem{DelducEta}
F.~Delduc, M.~Magro and B.~Vicedo,
``An integrable deformation of the AdS$_5$ x S$^5$ superstring action,''
Phys.\ Rev.\ Lett.\  {\bf 112}, no. 5, 051601 (2014), [arXiv:1309.5850[hep-th]];\\
``Derivation of the action and symmetries of the $q$-deformed $AdS_{5} \times S^{5}$ superstring,''
JHEP {\bf 1410}, 132 (2014), [arXiv:1406.6286[hep-th]].

\bibitem{eta}
G.~Arutyunov, R.~Borsato and S.~Frolov,
``S-matrix for strings on $\eta$-deformed AdS5 x S5,''
JHEP {\bf 1404}, 002 (2014)
[arXiv:1312.3542[hep-th]];\\
%
B.~Hoare, R.~Roiban and A.~A.~Tseytlin,
``On deformations of $AdS_n$ x $S^n$ supercosets,''
JHEP {\bf 1406}, 002 (2014)
[arXiv:1403.5517[hep-th]];\\
%
O.~Lunin, R.~Roiban and A.~A.~Tseytlin,
``Supergravity backgrounds for deformations of AdS$_{n} \times S^n$ supercoset string models,''
Nucl.\ Phys.\ B {\bf 891}, 106 (2015), [arXiv:1411.1066[hep-th]];
\\
%
B.~Hoare,
``Towards a two-parameter q-deformation of AdS$_3 \times S^3 \times M^4$ superstrings,''
Nucl.\ Phys.\ B {\bf 891}, 259 (2015), [arXiv:1411.1266[hep-th]];\\
%
S.~J.~van Tongeren,
``On classical Yang-Baxter based deformations of the AdS$_{5}$ x S$^{5}$ superstring,''
JHEP {\bf 1506}, 048 (2015), [arXiv:1504.05516[hep-th]];\\
G.~Arutyunov, R.~Borsato and S.~Frolov,   
``Puzzles of $\eta$-deformed AdS$_5 \times$ S$^5$,''
JHEP {\bf 1512}, 049 (2015)
[arXiv:1507.04239[hep-th]];
%
\\
G.~Arutyunov, S.~Frolov, B.~Hoare, R.~Roiban and A.~A.~Tseytlin,
``Scale invariance of the $\eta$-deformed $AdS_5\times S^5$ superstring, T-duality and modified type II equations,''
Nucl.\ Phys.\ B {\bf 903}, 262 (2016),
[arXiv:1511.05795[hep-th]];\\
%
%
%
L.~Wulff and A.~A.~Tseytlin,
``Kappa-symmetry of superstring sigma model and generalized 10d supergravity equations,''
arXiv:1605.04884[hep-th];\\
R.~Borsato and L.~Wulff,
``Target space supergeometry of $\eta$ and $\lambda$-deformed strings,''
arXiv:1608.03570 [hep-th].




\bibitem{SftsGr} 
K.~Sfetsos,
``Integrable interpolations: From exact CFTs to non-Abelian T-duals,''
Nucl.\ Phys.\ B {\bf 880}, 225 (2014)
[arXiv:1312.4560[hep-th]].

\bibitem{MoreLmb} 
T.~J.~Hollowood, J.~L.~Miramontes and D.~M.~Schmidtt,
``Integrable Deformations of Strings on Symmetric Spaces,''
JHEP {\bf 1411}, 009 (2014), 
[arXiv:1407.2840[hep-th]];\\
%
T.~J.~Hollowood, J.~L.~Miramontes and D.~M.~Schmidtt,
``An Integrable Deformation of the $AdS_5 \times S^5$ Superstring,''
J.\ Phys.\ A {\bf 47}, no. 49, 495402 (2014)
[arXiv:1409.1538[hep-th]].
K.~Sfetsos and D.~C.~Thompson,
``Spacetimes for $\lambda$-deformations,''
JHEP {\bf 1412}, 164 (2014)
[arXiv:1410.1886[hep-th]];\\
%
C.~Appadu and T.~J.~Hollowood,
``Beta function of k deformed AdS$_{5}$ x S$^{5}$ string theory,''
JHEP {\bf 1511} (2015) 095, [arXiv:1507.05420[hep-th]];\\
%
B.~Hoare and A.~A.~Tseytlin,
``On integrable deformations of superstring sigma models related to $AdS_n \times S^n$ supercosets,''
Nucl.\ Phys.\ B {\bf 897}, 448 (2015)
[arXiv:1504.07213[hep-th]];\\
R.~Borsato, A.~A.~Tseytlin and L.~Wulff,
``Supergravity background of $\lambda$-deformed model for AdS$_2 \times$  S$^2$ supercoset,''
Nucl.\ Phys.\ B {\bf 905}, 264 (2016),
[arXiv:1601.08192[hep-th]];\\
Y.~Chervonyi and O.~Lunin,
``Supergravity background of the $\lambda$-deformed AdS$_3 \times$ S$^3$ supercoset,''
Nucl.\ Phys.\ B {\bf 910}, 685 (2016), [arXiv:1606.00394 [hep-th]];\\
R.~Borsato and L.~Wulff,
``Target space supergeometry of $\eta$ and $\lambda$-deformed strings,''
JHEP {\bf 1610}, 045 (2016), [arXiv:1608.03570 [hep-th]].

\bibitem{DSTLmbAdS5}
S.~Demulder, K.~Sfetsos and D.~C.~Thompson,
``Integrable $\lambda$-deformations: Squashing Coset CFTs and $AdS_5\times S^5$,''
JHEP {\bf 1507}, 019 (2015), [arXiv:1504.02781[hep-th]].


\bibitem{GenLambda1}
K.~Sfetsos, K.~Siampos and D.~C.~Thompson,
``Generalised integrable $\lambda$-- and $\eta$-deformations and their relation,''
Nucl.\ Phys.\ B {\bf 899}, 489 (2015), [arXiv:1506.05784 [hep-th]];\\
  Y.~Chervonyi and O.~Lunin,
``Generalized $\lambda$-deformations of AdS$_p \times$ S$^p$,''
Nucl.\ Phys.\ B {\bf 913}, 912 (2016),
[arXiv:1608.06641 [hep-th]];

\bibitem{GenLambda1.5}
O.~Lunin and W.~Tian,
``Analytical structure of the generalized $\lambda$-deformation,''
Nucl.\ Phys.\ B {\bf 929}, 330 (2018)
[arXiv:1711.02735 [hep-th]].

\bibitem{GenLambda2}
 G.~Georgiou and K.~Sfetsos,
``A new class of integrable deformations of CFTs,''
JHEP {\bf 1703}, 083 (2017)
[arXiv:1612.05012 [hep-th]];\\
  G.~Georgiou and K.~Sfetsos,
``The most general $\lambda$-deformation of CFTs and integrability,''
JHEP {\bf 1903}, 094 (2019)
[arXiv:1812.04033 [hep-th]];\\
  G.~Georgiou and K.~Sfetsos,
``Integrable Lorentz-breaking deformations and RG flows,''
arXiv:1902.05407 [hep-th].

\bibitem{WZW}
E.~Witten,
``Nonabelian Bosonization in Two-Dimensions,''
Commun.\ Math.\ Phys.\  {\bf 92}, 455 (1984).

\bibitem{PCM}
A.~M.~Polyakov,
``Interaction of Goldstone Particles in Two-Dimensions. Applications to Ferromagnets and Massive Yang-Mills Fields,''
Phys.\ Lett.\  {\bf 59B}, 79 (1975).

\bibitem{Duality}
R.~Dijkgraaf, H.~L.~Verlinde and E.~P.~Verlinde,
``String propagation in a black hole geometry,''
Nucl.\ Phys.\ B {\bf 371}, 269 (1992);\\
 A.~Giveon,
``Target space duality and stringy black holes,''
Mod.\ Phys.\ Lett.\ A {\bf 6}, 2843 (1991);\\
 E.~B.~Kiritsis,
``Duality in gauged WZW models,''
Mod.\ Phys.\ Lett.\ A {\bf 6}, 2871 (1991);\\
M.~Rocek and E.~P.~Verlinde,
``Duality, quotients, and currents,''
Nucl.\ Phys.\ B {\bf 373}, 630 (1992)
[hep-th/9110053].

\bibitem{Grigoriev:2007bu}
M.~Grigoriev and A.~A.~Tseytlin,
``Pohlmeyer reduction of AdS(5) x S**5 superstring sigma model,''
Nucl.\ Phys.\ B {\bf 800}, 450 (2008)
[arXiv:0711.0155 [hep-th]].

\bibitem{Lunin:2018vsn}
O.~Lunin and W.~Tian,
``Scalar fields on $\lambda$-deformed cosets,''
Nucl.\ Phys.\ B {\bf 938}, 671 (2019)
[arXiv:1808.02971 [hep-th]].


\bibitem{Ganor:2002ya}
O.~Ganor, M.~B.~Halpern, C.~Helfgott and N.~A.~Obers,
``The Outer automorphic WZW orbifolds on so(2n), including five triality orbifolds on so(8),''
JHEP {\bf 0212}, 019 (2002)
[hep-th/0211003].


\bibitem{Fradkin:1991ie} 
E.~S.~Fradkin and V.~Y.~Linetsky,
``On space-time interpretation of the coset models in D $<$ 26 critical string theory,''
Phys.\ Lett.\ B {\bf 277}, 73 (1992).


\bibitem{Bars:1991zt}
I.~Bars and K.~Sfetsos,
``A Superstring theory in four curved space-time dimensions,''
Phys.\ Lett.\ B {\bf 277}, 269 (1992)
[hep-th/9111040].











\end{thebibliography}
\end{document}